\documentclass[preprint]{revtex4}
\usepackage{amsfonts,amsmath,amssymb}
\usepackage{hyperref}
\newcommand{\be}{\begin{equation}}
\newcommand{\ee}{\end{equation}}

\newcommand{\bea}{\begin{eqnarray}}
\newcommand{\eea}{\end{eqnarray}}

\newcommand{\nn}{\nonumber}

\newcommand{\ra}{\longrightarrow}
\newcommand{\cN}{{\cal N}}
\newcommand{\cA}{{\cal A}}
\newcommand{\Tr}{\mbox{Tr}}
\newcommand{\cZ}{{\cal Z}}
\begin{document}
\title{Bagger-Lambert theory on an orbifold \\ and its relation to Chern-Simons-matter theories}
\author{Nakwoo Kim}
\affiliation{Department of Physics and Research Institute of Basic Science, \\
Kyung Hee University, Seoul 130-701, Korea }
\email{nkim@khu.ac.kr}
%
%
%
\begin{abstract}
We consider how to take an orbifold reduction for
 the multiple M2-brane theory recently proposed by Bagger and Lambert, and discuss its relation to 
Chern-Simons theories.
Starting from the infinite dimensional 3-algebra realized as the Nambu bracket on a 3-torus,
we first suggest an orbifolding prescription for various fields. Then we introduce a 
second truncation, which effectively reduces the internal space to a 2-torus. Eventually one obtains 
a large-$N$ limit of Chern-Simons gauge theories coupled to matter fields. 
We consider an abelian orbifold $\mathbb{C}^4/\mathbb{Z}_n$,
and illustrate how one can arrive at the $\cN=6$ supersymmetric theories with gauge groups
$U(N)\times U(N)$ and Chern-Simons levels $(k,-k)$, as recently constructed by Aharony, Bergman, Jafferis and Maldacena.
\end{abstract}
\maketitle

\section{Introduction}
M-theory is conjectured to exist at certain limits in the moduli space of string theory, for instance as the strong coupling limit of IIA 
string theory. In spite of the extensive research activities undertaken since mid-90's, our knowledge is still rather limited: 
at low energy M-theory can be approximated by $D=11$ supergravity, and the fundamental degrees of freedom are probably M2-branes, 
and there are also  the electromagnetic dual partner M5-branes. Although there exist several non-perturbative, semi-microscopic descriptions, such as Matrix theory \cite{Banks:1996vh} or via AdS/CFT correspondence \cite{Maldacena:1997re}, one usually has to invoke string duality to study M-theory dynamics. In particular, for M-branes the action has been constructed for a single brane only, and how to incorporate interactions remained an impending problem.

Recently a very interesting result is published by Bagger and Lambert, who succeeded in constructing
 a new, maximally superconformal three dimensional field theory actions which arguably describe multiple M2-branes
 \cite{Bagger:2006sk,Bagger:2007jr,Bagger:2007vi}.
 The Bagger-Lambert (BL) theory is based on a 3-algebra structure in contrast to the conventional Lie-algebra. The totally antisymmetric 3-product has been called for in order to describe the Myers effect \cite{Myers:1999ps} which predicts that multiple M2-branes in external field should expand into a fuzzy 3-sphere, to make M5-branes. The BPS equations, or the generalized Nahm equations are presented and studied earlier in \cite{Basu:2004ed}, and the 3-algebra structures are also independently explored in \cite{Gustavsson:2007vu,Gustavsson:2008dy}.

Although the BL construction provides a very elaborate setup relevant for multiple M2-branes, it is soon realized that when the 3-algebra structure is to be combined with the maximal supersymmetry, there is basically only one gauge group allowed: $SO(4)$, when one demands as usual a finite dimensional vector space with positive-definite norm \cite{nogo}. Apparently there then remain two avenues to pursue, if one sticks to maximal supersymmetry. One might try to make sense of the BL theory with an infinite dimensional gauge group, or one starts to consider the case where the metric of the vector space has a Lorentzian signature. Containing ghost fields at classical level, the latter possibilities might suffer from the lack of unitarity. A concrete construction which can incorporate any conventional Lie-group is given and claimed to be unitary, but it seems to be more intimately related to the maximally supersymmetric Yang-Mills theory rather than giving rise to genuinely new theories. For the works along this direction see for instance \cite{lorentzian}.

It thus seems that the maximal supersymmetry for multiple M2-branes is extremely constraining. 
On the other hand, similar Chern-Simons gauge theories with less, yet extended supersymmetries with arbitrary number of fields can be 
constructed. $\cN=4$ theories are constructed, originally using IIB construction, in  \cite{Gaiotto:2008sd}, see also \cite{Hosomichi:2008jd}. 
More recently Chern-Simons-matter systems with $\cN=6$ are given by Aharony, Bergman, Jafferis and Maldacena (ABJM) \cite{Aharony:2008ug}, 
as multiple M2-branes action probing an orbifold singularity $\mathbb{C}^4/\mathbb{Z}_k$. 
The order of orbifold group $k$
appears as the level of Chern-Simons terms in the dual field theory. Unlike the maximally supersymmetric case, these theories come in an infinite family of actions, with the familiar quiver gauge theory pattern. As it is elegantly summarised in \cite{Gaiotto:2008sd}, for $\cN\ge 4$ the quiver data are succinctly encoded in super-Lie algebras. There are more recent works which explored the precise mathematical conditions for $\cN=5,6$ supersymmetry. Readers are referred to \cite{Hosomichi:2008jb,Bagger:2008se,Schnabl:2008wj} also.

In this paper we choose to study the Bagger-Lambert theory with an infinite dimensional vector space. An advantageous way of viewing it is as a gauge theory of volume-preserving diffeomorphisms on a three dimensional space $\Sigma$ endowed with a metric, as described for instance in Sec.~6 of \cite{Bagger:2007vi}. The 3-product is simply given as the Nambu bracket. The no-go theorem 
proved in \cite{nogo} asserts that it is out of the question to consistently truncate the infinite dimensional vector space 
to a finite dimensional one (except for $SO(4)$) with a 3-algebra satisfying the so-called fundamental identity 
which is required for supersymmetry and gauge invariance. 

Instead, we take the Nambu-bracket theory as it is and propose an orbifold prescription on it.  
Once we perform the projection and break some of the supersymmetries, 
one can identify a second consistent truncation after which the theory can be seen as a large-$N$
limit of conventional Chern-Simons-matter theories. This second step essentially reduces the internal space into a 2-torus, by fixing the Fourier 
index along one of the internal space dimensions. Then the resulting theory can
 be expressed in terms of Poisson brackets, which is amenable to matrix regularization.

The second truncation is compatible with the given orbifold: it can be understood as demanding 
invariance for all different orders of the orbifold group, so the supersymmetry is not broken, except
for the special cases $n=1,2$.
In this paper we consider orbifold $\mathbb{C}^4/\mathbb{Z}_n$. After the first step, we have 
$\cN=6$ (if $n>2$) theory with an infinite dimensional 3-algebra structure. Our observation is that 
for all $n$ there is a universal $\cN=6$ subsector of the orbifolded BL theory, and this subsector 
is a large-$N$ limit of the ABJM model. The regularised theory exhibits $U(N)\times U(N)$ gauge symmetry 
with Chern-Simons level
$k=4\pi^2/\kappa N^2$, where $\kappa$ is the coupling constant of
 3-algebra theory. Quantum mechanically, $k$ should take integer values and the vacuum
moduli space of the ABJM model is of course (a symmetric product of) $\mathbb{C}^4/\mathbb{Z}_k$.

The study of multiple M2-branes following the Bagger-Lambert proposal attracted a tremendous amount of interest recently. 
A list of papers are given in \cite{others}.

This paper is organized as follows. In Sec.~2 we briefly review the Bagger-Lambert action mainly to setup the notations. 
We also introduce the 3-algebra defined as a volume-preserving diffeomorphisms on the 3-torus. 
In Sec.~3 we explain how the orbifold projection can be taken on the Bagger-Lambert 3-algebra action, 
and in particular by considering $\mathbb{C}^4/\mathbb{Z}_n$ orbifolds we argue one can obtain 
the ABJM models with $U(N)\times U(N)$ gauge group. In Sec.~4 we conclude with discussions.

\section{Review of the Bagger-Lambert theory and the large-N limit}
In the BL model the matter fields take values in a vector space $\cA$ which is endowed with a 3-product. 
If we introduce basis vectors $T^a$, the structure constants are given from
\be
[ T^a , T^b , T^c ] = f^{abc}_{\;\;\;\;\;\;d} T^d
\ee
We assume that the vector space is given a positive-definite metric $h^{ab}$, which can be used to raise or lower the indices. 
The inner-product will be called $\Tr$. We also demand the following {\it invariance} property,
\bea
\Tr ( [ T^a , T^b , T^c ] , T^d ) = -
\Tr ( T^a , [ T^b , T^c , T^d ]  )
\eea
which implies $f^{abcd}$ is totally antisymmetric. We also need the {\it fundamental} identity,
\be
[T^a, T^b , [ T^c , T^d , T^e ]]
=
[[ T^a , T^b , T^c ] , T^d , T^e]
+
[T^c , [ T^a , T^b , T^d ] , T^e ]
+
[ T^c , T^d , [T^a , T^b , T^e ]]
\ee
The gauge transformation is expressed in terms of 3-algebra. One is requested to pick up two elements in the vector space to depict an infinitesimal gauge transformation. For instance, with $\alpha,\beta\in \cA$,
\be
\delta X = [ \alpha, \beta , X ] ,  \;\; \mbox{or} \;\;
\delta X_d = f^{abc}_{\;\;\;\;\;\; d} \alpha_a \beta_b X_c .
\ee
The Lagrangian density for BL theory is given as follows,
\bea
{\cal L} &=&
-\frac{1}{2} D_\mu X^{aI} D^{\mu} X^I_a
+
\frac{i}{2} \bar{\Psi}^a \Gamma^\mu D_\mu \Psi_a
+
\frac{i}{4} \bar{\Psi}_b\Gamma_{IJ} X^I_c X^J_d \Psi_a f^{abcd}
\nn\\
&&
- V(X)
+
\frac{1}{2} \varepsilon^{\mu\nu\lambda}
(f^{abcd} A_{\mu ab} \partial_\nu A_{\lambda cd} +
\frac{2}{3} f^{cda}_{\;\;\;\;\;\; g} f^{efgb} A_{\mu ab}
A_{\nu cd} A_{\lambda e f}
)
.
\eea
where $I=1,2,\cdots 8$ and $X^I$ represent the location of M2-branes in the eight dimensional transverse
space. $\Psi$ is the superpartner fermion which come in spinor representation of $SO(8)$. $\Gamma_\mu , ( \mu=0,9,10)$ and $\Gamma^I$ are the 11 dimensional gamma matrices in the Majorana representation. The scalar potential is given in terms of the square of 3-algebra products,
\be
V = \sum_{I,J,K=1}^{8} \frac{1}{12} \Tr
(
[ X^I, X^J , X^K]
[X^I, X^J, X^K ]
).
\label{sp}
\ee
In \cite{Bagger:2007jr} it is shown that the above Lagrangian is invariant under the following 16 supersymmetry transformatons:
\bea
\delta X^I_a
&=&
i \bar{\epsilon} \Gamma^I \Psi_a
\\
\delta \Psi_a
&=&
D_\mu X^I_a \Gamma^\mu \Gamma^I \epsilon
 - \frac{1}{6} X^I_b X^J_c X^K_d f^{bcd}_{\;\;\;\;\;\; a} \Gamma^{IJK} \epsilon
\\
\delta \tilde{A}^{\;\;b}_{\mu\;\; a} &=&
i \bar{\epsilon} \Gamma_\mu \Gamma_I X^I_c \Psi_d f^{cdb}_{\;\;\;\;\;\; a}
\label{susy_vec}
\eea
The covariant derivatives are chosen as
\be
D_\mu X_a = \partial_\mu X_a - \tilde{A}^{\;\;b}_{\mu\;\; a} X_b
\ee
with $\tilde{A}^{\;\;b}_{\mu\;\; a} = f^{cdb}_{\;\;\;\;\;\; a} A_{\mu cd}$.
More details can be found in \cite{Bagger:2007jr}.

An infinite dimensional realization of the 3-algebra structure is given by the space of differentiable functions on a 3-manifold $\Sigma$. 
In concreteness when $X,Y,Z$ are functions on $\Sigma$, with local coordinates $\sigma_i, i=1,2,3$ and metric $g$,
\be
[ X , Y , Z ] := \frac{1}{\sqrt{g}}
\varepsilon^{ijk} \partial_i X \partial_j Y\partial_k Z
\ee
One can check that the 3-algebra defined as above satisfies the invariance property and fundamental identity. In Ref.~\cite{Bagger:2007vi} the 3-sphere was chosen for $\Sigma$ and it is used to describe the fuzzy sphere solution.

In this paper we choose for simplicity the three dimensional torus $T^3$ instead, 
for $\Sigma$. $\sigma_i$ are the angular coordinates with periodicity $2\pi$. 
It is convenient to assume that the vectors $X,Y$ take complex values.
Obviously we can define the inner product as follows,
\be
(X, Y) \equiv
\Tr ( X^* Y )
= \frac{1}{(2\pi)^3}\int d^3\sigma X^* Y . 
\ee
The basis vectors are simply given by the Fourier modes, and labeled by 3-vectors in $\mathbb{Z}^3$.
\be
X(\sigma_1, \sigma_2, \sigma_3) = \sum_{\vec{N}\in \mathbb{Z}^3} X_{\vec{N}} \, e^{i \vec{N} \cdot \vec{\sigma}}
\ee
It is also a simple matter to evaluate the structure constants,
\be
f_{\vec{N}_1, \vec{N}_2 , \vec{N}_3 , \vec{N}_4} = i \kappa ( \vec{N}_1 \times \vec{N}_2 \cdot \vec{N}_3 ) \,
\delta ( \vec{N}_1+ \vec{N}_2 + \vec{N}_3 + \vec{N}_4 ) ,
\ee
where the Kronecker delta $\delta(\vec{N})$ gives the value $1$ when $\vec{N}$ is zero, otherwise it is $0$. $\kappa$ is a free
parameter which plays the role of a coupling constant. 

\section{Taking an orbifold  $\mathbb{C}^4/\mathbb{Z}_n$ of 3-algebra}
As a general statement, when the orbifold action makes a discrete subgroup of $SU(4)\subset SO(8)$, 
the supersymmetry of M2-branes in that background is broken either to $\cN=2$ or $\cN=0$, 
depending on the orientation. We will always assume that the M2-branes have the right orientation, 
so in general the orbifolded M2-brane theories have $\cN=2$ supersymmetry. 

There are however special cases which exhibit enhanced supersymmetry. 
Let us first define the complex coordinates of $\mathbb{C}^4$ in the following way.
\be
Z^A = (X^{2A-1} + i X^{2A})/\sqrt{2}, \quad A=1,2,3,4.
\ee
It will be convenient sometimes to use an alternative complexification,
\be
\tilde{Z}^1 = Z^1, \quad \tilde{Z}^2=Z^2, \quad \tilde{Z}^3 = Z^{*3},
\quad \tilde{Z}^4 = Z^{*4} .
\ee
The discrete symmetry group of our interest is generated by the following action
\be
\tilde{Z}^A \ra \omega \tilde{Z}^A, \quad A=1,2,3,4.
\ee
where $\omega=\exp{(2\pi i/n)}$. 
Since for $Z^A$ the transformation matrix is given as $\mbox{diag}(\omega,\omega,\omega^{-1},\omega^{-1})\in SU(4)$ this defines 
a supersymmetric orbifold. It turns out that in fact the unbroken supersymmetry is $\cN=6$, 
or $3/4$ for M2-branes of the right orientation. 

The number of remaining supersymmetries can be enumerated as the number of
spinors left invariant under the rotation group given above. On a spinor, of course the rotation generators are given as
\be
R= \Gamma_{12}+\Gamma_{34}-\Gamma_{56}-\Gamma_{78} . 
\label{rot}
\ee
And we are presented with a simple question of counting the zero modes. This can be done as follows. 
We first re-express the above as $\Gamma_{12}(1-\Gamma_{1234}+\Gamma_{1256}+\Gamma_{1278}$). 
And we notice that $\Gamma_{1234},\Gamma_{1256},\Gamma_{1278}$ are all Hermitian and commute with each other, so they can be simultaneously diagonalized. Furthermore they are all traceless and have equal numbers of eigenvectors for eigenvalue $1$ and $-1$. In the chiral representation with $\Gamma_{12345678}=1$ they are eight dimensional, and one can easily convince oneself that in a simple basis they must be expressed as
\begin{center}
\begin{tabular}{|c|cccccccc|}
  \hline
  $\Gamma_{1234}$ & $+$ & $+$ & $+$ & $+$ & $-$ & $-$ & $-$ & $-$ \\
  $\Gamma_{1256}$  & $+$ & $+$ & $-$ & $-$ & $+$ & $+$ & $-$ & $-$ \\
  $\Gamma_{1278}$ & $-$ & $-$ & $+$ & $+$ & $+$ & $+$ & $-$ & $-$ \\
  \hline
\end{tabular}
\end{center}
Using this basis it is clear that 6 out of 8 basis vectors have eigenvalue zero for the operator $R$.
Namely, they are 1,2,3,4,6, and the 8-th entries.

Orbifolding implies that the points related by the discrete symmetry group should be identified. 
For the gauge theory on the branes, the fields should be identified, {\it up to a common global gauge transformation}.
This is a well-known procedure and has been extensively used for D-brane dynamics where the fields are in the adjoint representation of Lie algebras, 
so the fields evaluated at different points are to be identified up to a unitary transformation \cite{Douglas:1996sw}. In physical terms, the global gauge transformation relates the D-brane with its 
mirror images in the covering space.

For the infinite dimensional 3-algebras we are interested in, the gauge transformations are generated by a 3-product and 
we are asked to choose two elements of the vector space. As a simple choice let us consider the following gauge transformation.
\be
\delta X = [ \sigma_1 , \sigma_2 , X ] . 
\ee
Obviously this results in $\delta X = i \partial_3 X$ so the finite gauge transformation is given by {\it translations} along the coordinate 
$\sigma_3$. 
Now we are ready to give the orbifolding prescription for our 3-algebra. 
On a 3-algebra valued field $\tilde{Z}^A$, we propose to consider the following discrete symmetry
\be
\tilde{Z}^A (\sigma_1 , \sigma_2 , \sigma_3 ) \ra \omega \tilde{Z}^A (\sigma_1 , \sigma_2 , \sigma_3 -2\pi/n ),
\label{zktr}
\ee
i.e. we have combined the $\mathbb{Z}_n$ action on $\mathbb{R}^8$ with another $\mathbb{Z}_n$ discrete translation invariance 
in the auxiliary space $\Sigma$. 
Note that, were it not for the translation in the internal space, 
there would not be any invariant mode. It is clear that this is just one of many different 
possibilities, but we will see that this particular choice leads to the large $N$ limit
of ABJM model.   

Now if we impose invariance, the solutions are given simply as the Bloch waves,
\be
\tilde{Z}^A (\sigma_1 , \sigma_2 , \sigma_3 ) = \tilde{\cZ}^A (\sigma_1 , \sigma_2 , \sigma_3 ) \, e^{i\sigma_3}, 
\label{b1}
\ee
where $\tilde{\cZ}$ is periodic in $\sigma_3$ in the following way.
\be
\tilde{\cZ} (\sigma_1 , \sigma_2 , \sigma_3 )
=
\tilde{\cZ} (\sigma_1 , \sigma_2 , \sigma_3 + 2\pi/n) . 
\ee
For fermions, the orbifolding operations are required to be compatible with the action and
the supersymmetry. From the invariance of the Yukawa interaction, $\bar{\Psi}_a
\Gamma_{IJ} X^I_b X^J_c \Psi_d f^{abcd}$, we can easily see that for $\Psi$
the discrete symmetry transformation should be defined as
\be
\Psi(\sigma_1,\sigma_2,\sigma_3) \ra e^{-\pi iR/n}
\Psi(\sigma_1,\sigma_2,\sigma_3-2\pi/n)
\ee
where $R$ is the rotation generator defined in Eq.~\eqref{rot}.

Before considering the gauge fields $A^{ab}_\mu$, 
let us first consider the scalar potential Eq.~\eqref{sp} and try to re-express it in terms of the complex fields $Z^A$. 
The BL Lagrangian is maximally supersymmetric so in principle it can be written as an $\cN=2$ theory, i.e.  
using four chiral multiplets instead of eight real scalars and an $SO(8)$ spinor field. 
The interactions must be determined by a quartic superpotential. 
The superfield description is studied in \cite{Bagger:2006sk} and a superpotential was proposed.
It is given simply as
\bea
W &=&
\frac{1}{12} \Tr ( [Z^A, Z^B, Z^C] Z^D )
\nn\\
&=&   2\,  \Tr ( [Z^1 , Z^2 , Z^3 ] Z^4 ) . 
\eea
As usual, the scalar potential is given as the sum of F-term and D-term potentials.
After some algebra, one can verify that in terms of four complex scalar fields the potential function is given as follows.
\bea
V &=& \frac{1}{12} \Tr \sum_{I,J,K=1}^8
[ X^I, X^J, X^K
]^2
\nn\\
&=&
\sum_{A,B,C=1}^4
\Tr \left(
\frac{2}{3}
|
[
Z^A,Z^B,Z^C
]
|^2
+
\frac{1}{2}
[
Z^A,\bar{Z}^A,Z^C
]
[
Z^B,\bar{Z}^B,\bar{Z}^C
]
\right) , 
\label{csp}
\eea
where the fundamental identity is used to establish the identity.

The orbifolding procedure as considered above essentially introduces a gradation into the vector space of 3-algebra. 
First of all the vector space is divided into two subsets.  
One can treat them as dual vector spaces, a half satisfying Eq.(\ref{b1}), 
and the other half, or the dual, satisfying a similar condition given by the complex conjugate. 
More concretely if we denote $\tilde{\cA} = \{ \tilde{Z}^A : A=1,2,3,4 \}$ then they are decomposed as
\be
\tilde{\cA} = \bigoplus_{l\in \mathbb{Z}} \tilde{\cA}_{ln+1} , 
\ee
where the elements of $\tilde{\cA}_{j}$ have momentum $j$ along $\sigma_3$.

It is a crucial observation that in the expression for the superpotential and also
for the ordinary potential function one always encounters 3-products with
one entry in $\tilde{\cA}$ and other two entries from $\tilde{\cA}^*$, or the opposite. 
There is no appearance of a 3-product $[\tilde{Z}^1, \tilde{Z}^2, \tilde{Z}^3]$, for instance in Eq.(\ref{csp}). 
This is also true for Yukawa-terms and most importantly for supersymmetry transformation rules. What this fact signifies is obvious. The orbifold operation is made so that the interactions as well as the supersymmetry respect the truncation. In other words, the interactions preserve the gradation introduced by the orbifold prescription.

Now it is also obvious that one can further take a consistent truncation of the fields where we keep only a single subspace, 
say $\tilde{\cA}_1$ and its conjugate space. 
In summary, after orbifolding and a further consistent truncation we can restrict ourselves to the fields given 
now as a function of two variables $\sigma_1,\sigma_2$, instead of three variables $\sigma_i , i=1,2,3$. 
Then what do we have when we compute a 3-product? Let us take an example and compute $[Z^1,Z^2,Z^3]$. We can set
\bea
Z^1 = \cZ^1 (\sigma_1, \sigma_2 ) e^{+i\sigma_3},\quad\quad
Z^2 = \cZ^2 (\sigma_1, \sigma_2 ) e^{+i\sigma_3},
\nn\\
Z^3 = \cZ^3 (\sigma_1, \sigma_2 ) e^{-i\sigma_3},\quad\quad
Z^4 = \cZ^4 (\sigma_1, \sigma_2 ) e^{-i\sigma_3}.
\label{trunca}
\eea
Then we can easily see
\be
[Z^1 , Z^2 , Z^3 ]
= i \kappa \left(
\cZ^1 \{ \cZ^2, \cZ^3 \} + \cZ^2\{ \cZ^3 , \cZ^1 \} - \cZ^3\{ \cZ^1 , \cZ^2 \}
\right) e^{i\sigma_3} , 
\ee
where the curly brackets denote the Poisson bracket on $(\sigma_1,\sigma_2)$-plane, i.e.
\be
\{ f(\sigma_1,\sigma_2) , g(\sigma_1,\sigma_2) \}
=
\partial_1 f \partial_2 g - \partial_2 f \partial_1 g . 
\ee
Then this 3-product contributes to the potential as
$\Tr |[ Z^1 , Z^2 , Z^3]|^2$, and in our setup this becomes
\be
\frac{\kappa^2}{(2\pi)^2} \int d^2\sigma
\Big|
\cZ^1 \{ \cZ^2, \cZ^3 \} + \cZ^2\{ \cZ^3 , \cZ^1 \} - \cZ^3\{ \cZ^1 , \cZ^2 \}
\Big|^2 . 
\ee
Now one can choose to introduce a new 3-product in the space of functions in $T^2$. This new 3-product is not totally antisymmetric 
but antisymmetric only with the first two entries, and defined in the following way.
\bea
\langle \cZ^1 , \cZ^2 ; \cZ^3 \rangle
&\equiv &
i \kappa(
\cZ^1 \{ \cZ^2, \cZ^3 \} + \cZ^2\{ \cZ^3 , \cZ^1 \} - \cZ^3\{ \cZ^1 , \cZ^2 \}
) . 
\label{new3}
\eea
Then the scalar potentials can be expressed as a sum of squares of this new 3-product.

Now let us move to the gauge fields and look for the orbifolding procedure which is compatible with 
the rules we have chosen for $X,\Psi$. Unlike the matter fields, the gauge fields have double 3-algebra indices and we cannot
treat them as an ordinary function defined on $\Sigma$. One might instead consider 
\be
A^{\mu}(\vec{\sigma}_1,\vec{\sigma}_2) = \sum_{\vec{M},\vec{N}\in \mathbb{Z}^3}
 A^{\mu}_{\vec{M}\vec{N}} e^{ i ( \vec{M} \cdot \vec{\sigma}_1 +\vec{N} \cdot \vec{\sigma}_2 ) }  \;\; ,
\ee
i.e. treat $A^\mu$ as a function on $\Sigma\times \Sigma$. We however do not find this interpretation very illuminating, and decide to 
work with the component fields $A^{\mu}_{\vec{M}\vec{N}}$ instead. By construction they satisfy antisymmetric property, 
\be
A^\mu_{\vec{M}\vec{N}} = - 
A^\mu_{\vec{N}\vec{M}} .
\ee
Furthermore, we require that for a real scalar field $X(\vec{\sigma})$, satisfying
\be
X(\vec{\sigma}) = \sum X_{\vec{M}} e^{i\vec{M}\cdot\vec{\sigma}}, \quad 
X_{\vec{M}}=X^*_{-\vec{M}} , 
\ee its covariant derivatives should be also real-valued, so we
easily infer that 
\be
\tilde{A}^\mu_{\vec{M}\vec{N}} = 
\tilde{A}^{*\mu}_{-\vec{M},-\vec{N}} . 
\ee

We are now ready to look for the appropriate transformation rules for the vector fields $\tilde{A}$. If we 
express the $\mathbb{Z}_n$ transformation Eq.(\ref{zktr}) in terms of the component fields, we get 
\be
\tilde{Z}_{\vec{M}} \ra e^{ \frac{2\pi i }{n}(1-M_3)}\tilde{Z}_{\vec{M}}  , 
\ee
and invariance requires $M_3 = 1 \mbox{ mod } n$, as we know already. We again demand that taking the covariant
derivatives on a scalar field should respect this property, and find if we adopt 
\be
\tilde{A}^{\mu\vec{M}}_{\;\;\;\;\vec{N}}
\ra
e^{ \frac{2\pi i }{n}(N_3-M_3)}
\tilde{A}^{\mu\vec{M}}_{\;\;\;\;\vec{N}} , 
\label{dvec_tr}
\ee
then the requirement is fulfilled. Invariance under this discrete transformation of course implies that
\be
M_3+N_3 = 0 \mbox{ mod } n , \mbox{ for }
\tilde{A}^\mu_{\vec{M}\vec{N}} . 
\ee
 We then need to make sure if the above discrete transformation rule for the vector field leaves the Chern-Simons part of the action invariant.
 Unlike the covariant derivatives or supersymmetry transformation rules, the action itself is written in terms of the original gauge field
 $A^\mu_{\vec{M}\vec{N}}$. And for the infinite dimensional representation for 3-algebra at hand, it is not clear how to invert the relation 
 $\tilde{A}^{\mu}_{\vec{M}\vec{N}}=f^{\vec{P}\vec{Q}}_{\;\;\;\;\;\;\vec{M}\vec{N}} A^\mu_{\vec{P}\vec{Q}}$, so there is a subtlety. 
 We can however easily check that, if the discrete transformation for $A^\mu$ is defined in the same way as Eq.(\ref{dvec_tr}), i.e.
 \be
 A^{\mu\vec{M}}_{\;\;\;\;\vec{N}}
\ra
e^{ \frac{2\pi i }{n}(N_3-M_3)}
A^{\mu\vec{M}}_{\;\;\;\;\vec{N}} , 
 \ee
 then Eq.(\ref{dvec_tr}) immediately follows, and the Chern-Simons part of the action is also invariant under this transformation. 
 
 We have thus completely fixed how the orbifold action should be realized on various fields of the BL theory with an infinite dimensional
 3-algebra. For the matter fields we could take another truncation and fix the momentum along $\sigma_3$, and we would like to 
 extend this operation to vector fields in such a way that the truncation is compatible with the gauge and 
 supersymmetry transformation. From the supersymmetry transformation rule for $\tilde{A}^\mu_{\vec{M}\vec{N}}$ in Eq.(\ref{susy_vec}), we see 
 that one cannot demand  $M_3=N_3=\pm 1$ for 
 $\tilde{A}^{\mu\vec{M}}_{\;\;\;\;\;\vec{N}}$, because supersymmetry transformations in general would give rise to 
 components with different values of $M_3=N_3$ for 
 $\tilde{A}^{\mu\vec{M}}_{\;\;\;\;\;\vec{N}}$. We instead demand  $M_3=N_3=\pm 1$ for 
 ${A}^{\mu\vec{M}}_{\;\;\;\;\;\vec{N}}$, which is clearly consistent with the supersymmetry transformation rule.

Since with this second truncation we have fixed how the fields depend on $\sigma_3$, it seems as if the orbifold projection and the 
truncation we have taken effectively reduces the symmetry structure of three dimensional diffeomorphisms into a two dimensional one. 
The resulting action should be very similar to what would be obtained from a supermembrane action. 
As it is well-known, the infinite dimensional gauge theory of area-preserving diffeomorphism can be regularized to give an ordinary Yang-Mills theory. 
For references see for instance \cite{apd}, and also note that the relationship between the topology of the 2-surface and different Lie algebras 
has been used to describe orbifolds of Matrix theory in \cite{maorb}.

One can see that matrix algebra in the large-$N$ limit can realize the diffeomorphism algebra
 of $T^2$ in the following way. (See the references in \cite{apd} for the original construction.) 
 Here $N$, the size of the matrices, plays the role of a regulator. We set $\xi=e^{2\pi i/N}$ and 
 choose a pair of unitary and traceless matrices $U,V$, which are primitive $N$-th roots of unity.
\be
UV = \xi VU , \quad
U^N = V^N = 1 .
\ee
It then follows that a set of matrices defined as
\be
J^{\vec{n}} = \xi^{-\frac{n_1 n_2}{2}} U^{n_1} V^{n_2} , 
\ee
for $n_1,n_2=1,\cdots, N$ provides a basis for $N\times N$ Hermition matrices of $U(N)$.
We use lowercase letters to denote 2-vectors, like $\vec{n}=(n_1,n_2)$. $\vec{\sigma}$ 
should also denote a vector in $T^2$ when multiplied to a 2-vector, like $\vec{m}\cdot\vec{\sigma}$. Multiplication is easily computed, for instance
\be
J^{\vec{m}} J^{\vec{n}} = \xi^{\vec{m}\times \vec{n}/2} J^{\vec{m}+\vec{n}} , 
\ee
where $\vec{m}\times\vec{n}=m_1n_2-m_2n_1$. 
The commutator thus becomes
\be
[ J^{\vec{m}} ,J^{\vec{n}} ] = 2i\sin \left( \frac{\pi}{N} \vec{m}\times \vec{n} \right)
J^{\vec{m}+\vec{n}} . 
\ee
Since this commutator approaches
$ \frac{2\pi i}{N} (\vec{m}\times \vec{n}) J^{\vec{m}+\vec{n}} $ in the large $N$ limit, it is clear that up to a certain re-scaling 
the Poisson bracket
$\{ e^{i\vec{m}\cdot\vec{\sigma}} , e^{i\vec{n}\cdot\vec{\sigma}} \}=-e^{i (\vec{m}+\vec{n}) \cdot \vec{\sigma}}$ can be approximated by a matrix commutator.

Now let us consider the new 3-product defined above in Eq.(\ref{new3}). For basis vectors
$e^{i\vec{n}\cdot\vec{\sigma}}$ they are computed as follows.
\be
\langle
e^{i\vec{n}_1\cdot\vec{\sigma}},
e^{i\vec{n}_2\cdot\vec{\sigma}};
e^{i\vec{n}_3\cdot\vec{\sigma}}
\rangle
=
-i 
\kappa
 (\vec{n}_1\times \vec{n}_2 + \vec{n}_1 \times \vec{n}_3 + \vec{n}_3 \times \vec{n}_2 )
e^{i(\vec{n}_1 + \vec{n}_2 + \vec{n}_3 )\cdot \vec{\sigma}} .
\ee
Using the same mapping between $N\times N$ matrices and Fourier modes on $T^2$, 
one can identify the corresponding matrix 3-product in the following way. We consider
\bea
\langle
J^{\vec{n}_1},
J^{\vec{n}_2};
J^{\vec{n}_3}
\rangle
&\equiv&
\lambda
(
J^{\vec{n}_1}
J^{\vec{n}_3}
J^{\vec{n}_2}
-
J^{\vec{n}_2}
J^{\vec{n}_3}
J^{\vec{n}_1}
)
\nn\\
&=&
\lambda
\left(
\xi^{(\vec{n}_1\times \vec{n}_3+(\vec{n}_1+\vec{n}_3)\times \vec{n}_2)/2}
-
\xi^{(\vec{n}_2\times \vec{n}_3+(\vec{n}_2+\vec{n}_3)\times \vec{n}_1)/2}
\right)
J^{\vec{n}_1+\vec{n}_2+\vec{n}_3}
\nn\\
&=& 2 i \lambda \sin \left[
\frac{\pi(\vec{n}_1\times \vec{n}_2 + \vec{n}_1 \times \vec{n}_3 + \vec{n}_3 \times \vec{n}_2)}{N}
\right]
J^{\vec{n}_1+\vec{n}_2+\vec{n}_3} . 
\eea
So we can see that the matrix algebra again gives a realization of the new 3-algebra in the large-$N$ limit. More concretely, 
after the orbifold projection and truncation the scalar fields are written in terms of the mode expansion 
coefficients $\cZ=\sum_{\vec{m}}\cZ_{\vec{m}}e^{i\vec{m}\cdot\sigma}$. 
Alternatively we can use $J^{\vec{m}}/\sqrt{N}$ as normalized basis vectors. Then the above result tells us that, in the large $N$ limit,  
\bea 
\langle \cZ^1, \cZ^2 ; \cZ^3 \rangle_{\vec{m}}
= \frac{2\pi i\lambda }{N^2} 
\sum_{\vec{n}_1+\vec{n}_2+\vec{n}_3=\vec{m}} (
\vec{n}_1\times \vec{n}_2 + \vec{n}_1 \times \vec{n}_3 + \vec{n}_3 \times \vec{n}_2)
\cZ^1_{\vec{n}_1} \cZ^2_{\vec{n}_2} \cZ^3_{\vec{n}_3} .
\eea
And the two definitions agree if we set 
\be
\kappa = - \frac{2\pi}{N^2} \lambda . 
\ee

In fact this generalization of 3-product, relaxing total antisymmetry, has already appeared in the literature in relation with the 
ABJM models in \cite{Bagger:2008se}, where the total antisymmetric property of the structure constant for 3-algebra 
was relaxed from the beginning and a particular solution to the generalized fundamental identity has been found as
(see Eq.(52) of \cite{Bagger:2008se}.)
\bea
[
X,Y;\bar{Z}
]_{BL}&=& \lambda ( X Z^\dagger Y - Y Z^\dagger X )
\nn\\
&=& \langle X, Y ; Z^\dagger \rangle_{here} . 
\eea

In order to adapt to the notations used in \cite{Aharony:2008ug} or \cite{Bagger:2008se}, 
one has better first recall that the superpotential for the ABJM model is given in the same way as the familiar conifold theory, i.e.
\be
W =
\frac{2\pi}{k}
\Tr (A^a B_{\dot{c}} A^b B_{\dot{d}}) \epsilon_{ab}\epsilon^{\dot{c}\dot{d}} . 
\label{superpot}
\ee
The coupling constant $k$ is integral and appears also as the Chern-Simons level in the ABJM theory. 
In $\cN=2$ language the $SU(4)$ fundamental representation $Z^A$ are here decomposed into two $SU(2)$ doublets, 
$A^a,B_{\dot{b}}, a,\dot{b}=1,2$, which are respectively in $(N,\bar{N})$ and $(\bar{N},N)$ representation of $U(N)\times U(N)$. 
In terms of the complex scalar fields used in this paper so far, they are related as follows:
\bea
A^a &\longrightarrow& {\cZ}^a ,
\\
B_{\dot{b}} &\ra& {\cZ}^{\dot{b}+2} .
\eea
So the superpotential given in terms of the 3-product should become
\bea
W &=& 2 \, \Tr ( [ Z^1 , Z^2 , Z^3 ] Z^4 )
\nn\\
&=& 2 \, \Tr ( \langle \cZ^1 , \cZ^2 ; \cZ^3 \rangle \cZ^4 )
\nn\\
&\ra &
2 \lambda \, \mbox{Tr} (A_1B_1A_2B_2-A_1B_2A_2B_1) , 
\eea
provided the 3-algebra coupling constant is related to the Chern-Simons level of the ABJM model as
\be
\kappa= -\frac{4\pi^2}{k N^2} . 
\label{dic1}
\ee

We have just seen that a matrix regularization of the orbifolded BL theory reduces the superpotential 
of the theory written in terms of 3-algebra into the superpotential of the ABJM model which involves just 
ordinary matrix multiplications. We still need to check whether the truncation of gauge fields gives us
$U(N)\times U(N)$ gauge symmetry, and the matter fields are in bi-fundamental representations. Let us 
first introduce a two dimensional notation for gauge fields:
\be
A^\mu_{\vec{m}\vec{n}} = A^\mu_{\vec{M}\vec{N}}, 
\quad 
\mbox{with } 
\vec{M}=(\vec{m},1) ,\,\, 
 \vec{N}=(\vec{n},1)
 .
\ee

Then for a scalar field $Z_{\vec{M}}$ with $\vec{M}=(\vec{m},1)$, taking a covariant derivative gives 
\bea
D_\mu Z_{\vec{m}} &=& \partial_\mu Z_{\vec{m}} + 2 i \kappa 
\sum_{\vec{p}+\vec{q}+\vec{n}=\vec{m}}
[ \vec{p}\times\vec{q} - \vec{n}\times (\vec{p}+\vec{q}) ] 
A_{\mu\vec{p}\vec{q}} Z_{\vec{n}} .
\label{c_der}
\eea
Recall that the matrix regularization means we use matrices $J^{\vec{m}}\propto U^{m_1}V^{m_2}$ as basis vectors. 
Then the constraint $\vec{p}+\vec{q}+\vec{n}=\vec{m}$ above implies
we should identify the gauge field $A_{\vec{p}\vec{q}}$ in the expansion coefficient of $J^{\vec{p}+\vec{q}}$. It is then easy to see
that the gauge coupling term can be approximated as a linear combination of two matrix multiplication, as follows.
\be
A^L = i \alpha\sum_{\vec{p},\vec{q}} \xi^{+\vec{p}\times\vec{q}/2} 
A_{\vec{p}\vec{q}}
J^{\vec{p}+\vec{q}}, \quad
A^R = i \alpha\sum_{\vec{p},\vec{q}} \xi^{-\vec{p}\times\vec{q}/2} 
A_{\vec{p}\vec{q}}J^{\vec{p}+\vec{q}} ,
\ee
where $\alpha$ is a real-valued normalization constant.
Then in the large-$N$ limit we have 
\bea
  (A^L_{\mu} Z - Z A^R_{\mu})_{\vec{m}} 
 &=& 
i  \alpha\sum_{\vec{p}+\vec{q}+\vec{n}=\vec{m}}
\left(
\xi^{+\vec{p}\times\vec{q}/2} J^{\vec{p}+\vec{q}} J^{\vec{n}}
-
\xi^{-\vec{p}\times\vec{q}/2}  J^{\vec{n}} J^{\vec{p}+\vec{q}}
\right)  A_{{\mu}\vec{p}\vec{q}} Z_{\vec{n}}
\nn\\
&= &
-2\alpha
\sum_{\vec{p}+\vec{q}+\vec{n}=\vec{m}} \sin \left[
\frac{\pi(\vec{p}\times\vec{q}+(\vec{p}+\vec{q})\times\vec{n})}{N}
\right] J^{\vec{p}+\vec{q}+\vec{n}}
A_{{\mu}\vec{p}\vec{q}} Z_{\vec{n}}
\eea
It is obvious that the covariant derivative in Eq.(\ref{c_der}) can be reproduced using large matrices when we define the Yang-Mills symmetry so that (for $Z^1,Z^2$)
\be
D_{\mu} Z = \partial_{\mu} Z +i ( A^L_\mu Z - Z A^R_\mu),
\ee
and identify
\be
\alpha = \frac{4\pi}{kN} .
\label{dic2}
\ee
We can thus identify $Z^1,Z^2$ as $A^1,A^2$, and $Z^3,Z^4$ as $B_1,B_2$ in Eq.(\ref{superpot}).

From the matching of the superpotential and gauge coupling, it should naturally follow that the scalar potential and Yukawa terms
 agree between the 3-algebra theory and the ABJM model. Let us now consider the Chern-Simons part of the action. 
 In the Bagger-Lambert theory, the Chern-Simons-type kinetic terms are given as 
\be
\frac{1}{2} \varepsilon^{\mu\nu\lambda}
(f^{abcd} A_{\mu ab}  \partial_{\nu} A_{\lambda cd} +
\frac{2}{3} f^{cda}_{\;\;\;\;\;\; g} f^{efgb} A_{\mu ab}
A_{\nu cd} A_{\lambda e f}
).
\label{3cs}
\ee
And in the ABJM theory we have 
\be
-\frac{k}{4\pi} \varepsilon^{\mu\nu\lambda} \mbox{Tr} ( A^L_{\mu} \partial_\nu^{} A^L_\lambda +
\frac{2i}{3} A^L_\mu A^L_\nu A^L_\lambda 
- A^R_{\mu} \partial_\nu^{} A^R_\lambda -
\frac{2i}{3} A^R_\mu A^R_\nu A^R_\lambda ).
\label{ABJMcs}
\ee
One can easily see that in components the first term in Eq.(\ref{3cs}) becomes 
\be
-2i \kappa \varepsilon_{\mu\nu\lambda}
\sum_{\vec{m}+\vec{n}+\vec{p}+\vec{q}=0} (\vec{m}\times \vec{n} + (\vec{m}+\vec{n})\times \vec{p} )
A^{\mu}_{\vec{m}\vec{n}}
\partial^\nu 
 A^{\lambda}_{\vec{p}\vec{q}} . 
\ee
The same expression is obtained from the ABJM model, 
\bea
\lefteqn{-\frac{k}{4\pi} \varepsilon^{\mu\nu\lambda} \mbox{Tr} ( A^L_{\mu} \partial_\nu^{} A^L_\lambda - A^R_{\mu} \partial_\nu^{} A^R_\lambda )}
\\
&& = \frac{ik\alpha^2 }{2} \varepsilon_{\mu\nu\lambda}
\sum_{\vec{m}+\vec{n}+\vec{p}+\vec{q}=0}
(\vec{m}\times \vec{n} + (\vec{m}+\vec{n})\times \vec{p} )
A^{\mu}_{\vec{m}\vec{n}}
\partial^\nu  A^{\lambda}_{\vec{p}\vec{q}} . 
\eea 
So they agree if
\be
\kappa = - \frac{k\alpha^2}{4} , 
\ee 
which is consistent with Eq.(\ref{dic1}) and Eq.(\ref{dic2}). 

When it comes to the cubic coupling term, one can show that the second term in Eq.(\ref{3cs}) can be simplified to 
\be
4\kappa \varepsilon^{\mu\nu\lambda}
 \sum_{\vec{m}+\vec{n}+\vec{p}+\vec{q}+\vec{r}+\vec{s}=0}
A_{\mu\vec{m}\vec{n}} A_{\nu\vec{p}\vec{q}} A_{\lambda\vec{r}\vec{s}} \, 
(\vec{m}\times\vec{n})(\vec{p}+\vec{q})\times(\vec{r}+\vec{s}) ,
\ee
and the cubic term in Eq.(\ref{ABJMcs}) gives exactly the same 
expression, if we again demand Eq.(\ref{dic1}) and Eq.(\ref{dic2}). 

We have thus established that the large $N$ limit of the ABJM model can provide a discretized version of orbifolded
BL model. 

\section{Discussions}
In this paper we discussed how one can take an orbifold projection on the 3-algebra of volume-preserving 
diffeomorphisms which gives an infinite dimensional realization of the Bagger-Lambert theory. 
We find it very interesting to see how the orbifolding procedure provides a natural method
to reduce the internal three-dimensional structure down to two dimensions. To obtain a sensible 
theory, one has to combine the discrete transformation in $\mathbb{R}^8$ with some transformation 
in the auxiliary $T^3$. We have made the simplest choice in Eq.(\ref{zktr}), and the resulting 
action can be matrix-regularised to give the ABJM model.


We know that 
quantum mechanically the coupling constant of the Chern-Simons action should be quantized.
For the BL theory with $SO(4)$ gauge group or the ABJM model, 
the residual symmetry under large gauge transformations dictates us to interpret the Chern-Simons level as the order of the orbifold
group, 
through a careful analysis of the vacuum moduli space \cite{0804.1114,Aharony:2008ug}. For an infinite dimensional 3-algebra, the BL theory
is not an ordinary quantum field theory because it has an infinite number of fields. 
So we would rather treat the BL theory only as a classical action in this article.

We studied the orbifold $\mathbb{C}^4/\mathbb{Z}_n$ in this paper, but in fact the resulting action from the 3-algebra theory 
does not contain the parameter $n$ as it stands.
It is because of the second truncation we have taken in Eq.(\ref{trunca}). At this point the order of orbifold
$n$ simply disappears from the theory since we restrict the mode number along $\sigma_3$ to 1. When we allow a tunable
coupling constant $\kappa$ for 3-algebra, it can be related to the combination $kN^2$ in the $U(N)\times U(N)$ gauge theory, 
as in Eq.(\ref{dic1}). Once we obtain ordinary Chern-Simons-matter theory with finite ranks, one
 can argue $k$ should be integer-valued,
and the vacuum moduli space is $\mbox{Sym}_N (\mathbb{C}^4/\mathbb{Z}_k)$.

One might ask whether we can somehow witness the supersymmetry enhancement for the special case of $\mathbb{C}^4/\mathbb{Z}_2$, 
in our procedure. It is well known that for this particular orbifold the supersymmetry is intact \cite{Aharony:2008ug}. 
 This means we lose no supersymmetry 
when we restrict the mode number along $\sigma_3$ to odd integers. But when we restrict it to $\pm 1$, we are discarding 
the information on the specific value of $n$, and the supersymmetry is reduced to $\cN=6$.

We admit the physical meaning of the Bagger-Lambert theory with infinite-dimensional 3-algebra is still rather obscure, so do not claim that it can be directly related to 
M-theory. But our work illustrates that one can at least take the BL theory as a tool to generate supersymmetric
Chern-Simons-matter theories relevant to M2-branes. It would be nice if we can incorporate stringy features of orbifolding procedure, 
such as discrete torsion and fractional branes.

Although for simplicity we have considered only the simplest orbifold $\mathbb{C}^4/\mathbb{Z}_n$ in this paper, 
obviously one can consider other, probably less supersymmetric, orbifolds. According to a systematic field theoretical 
analysis \cite{Hosomichi:2008jb,Schnabl:2008wj} there exist for instance $\cN=6$ theories with $U(1)\times USp(2N)$ gauge group. 
It is not clear how one can get such theories through orbifolding. Further exploration on orbifolds of 3-algebra is certainly very desirable.

\acknowledgments
We thank Sangmin Lee, Sungjay Lee and especially Jeong-Hyuck Park for discussions.
This research is supported by the Science Research Center Program of the Korea Science and Engineering Foundation through the Center for Quantum Spacetime (CQUeST) of Sogang University with grant number R11-2005-021, and also partly by the Korea Research Foundation Grant No. KRF-2007-331-C00072.

\end{document}